\documentclass[aps,prl,twocolumn,showpacs,preprintnumbers,amsmath,amssymb,superscriptaddress,10pt,floats]{revtex4-1}

\usepackage{graphicx}
\usepackage{dcolumn}
\usepackage{color}
\allowdisplaybreaks

\newcommand{\bvec}[1]{\ensuremath{\mathbf{#1}}}

\begin{document}
\title{Long-range Coulomb interactions in surface systems: a first principles description within self-consistently combined GW and dynamical mean field theory}

\author{P.~Hansmann}
\affiliation{Centre de Physique Th{\'e}orique, Ecole Polytechnique, CNRS-UMR7644, 91128 Palaiseau, France}

\author{T.~Ayral}
\affiliation{Centre de Physique Th{\'e}orique, Ecole Polytechnique, CNRS-UMR7644, 91128 Palaiseau, France}
\affiliation{Institut de Physique Th\'eorique (IPhT), CEA, CNRS, URA 2306, 91191 Gif-sur-Yvette, France} 

\author{L.~Vaugier}
\affiliation{Centre de Physique Th{\'e}orique, Ecole Polytechnique, CNRS-UMR7644, 91128 Palaiseau, France}

\author{P.~Werner}
\affiliation{Department of Physics, University of Fribourg, 1700 Fribourg, Switzerland}

\author{S.~Biermann}
\affiliation{Centre de Physique Th{\'e}orique, Ecole Polytechnique, CNRS-UMR7644, 91128 Palaiseau, France}
\affiliation{Japan Science and Technology Agency, CREST, Kawaguchi 332-0012, Japan}
 

\begin{abstract}
Systems of adatoms on semiconductor surfaces display competing ground states and exotic spectral properties typical of two-dimensional correlated electron materials which are dominated by a complex interplay of spin and charge degrees of freedom. We report a fully \emph{ab initio} derivation of low energy Hamiltonians for the adatom systems Si(111):X, with X=Sn, Si, C, Pb, that we solve within self-consistent combined GW and dynamical mean field theory (``GW+DMFT''). Calculated photoemission spectra are in agreement with available experimental data. We rationalize experimentally observed tendencies from Mott physics towards charge-ordering along the series as resulting from substantial long-range interactions.
\end{abstract}

\date{\today}
\maketitle

Understanding the electronic properties of materials with strong electronic Coulomb correlations remains one of the biggest challenges of modern condensed matter physics. The interplay of delocalization and interactions is not only at the origin of exotic ground states, but also determines the excitation spectra of correlated materials. The ``standard model'' of correlated fermions, the Hubbard model, in principle captures these phenomena. Yet, relating the model to the material on a microscopic footing remains a challenge. Even more importantly, the approximation
of purely local Coulomb interactions can become severe in realistic materials, where long-range interactions and charge fluctuation physics cannot be neglected.

Systems of adatoms on semiconducting surfaces, such as Si(111):X with X=Sn, C, Si, Pb, have been suggested \cite{tosatti74} to be good candidates for observing low-dimensional correlated physics.
Commonly considered to be realizations of the one-band Hubbard model and toy systems for investigating many-body physics on the triangular lattice, such surfaces have been explored experimentally 
\cite{uhrberg85,grehk93,weitering93,carpinelli96,carpinelli97,weitering97,slezak99,lay01,lobo03,dudr04,pignedoli04,upton05,modesti07,cardenas09,zhang10,tournier11,cleassen11} and theoretically \cite{kaxiras90,brommer92,santoro99,hellberg99,aizawa99,profeta00,shi02,shi04,profeta05,profeta07,schuwalow10,chaput11,li11}.These so-called $\alpha$-phases show a remarkable variety of interesting physics including commensurate charge density wave (CDW) states \cite{carpinelli96,carpinelli97,lay01} and isostructural metal to insulator transitions (MIT)\cite{modesti07}. However, while specific systems and/or phenomena have been investigated also theoretically, a comprehensive understanding including materials trends is still lacking. A central goal of our work is to present a unified picture that relates, within a single framework, different materials (adatom systems), placing them in a common phase diagram.

We derive low-energy effective Hamiltonians \emph{ab initio} from a combined density functional and constrained random phase approximation (cRPA) scheme \cite{aryasetiawan04} in the implementation of \cite{vaugier12} (see also the extension to surface systems in \cite{jpcm12}). While the first surprise are the relatively large values of the onsite interactions which we find to be of the order of the bandwidth ($\approx 1$ eV), most importantly we show that non-local interactions are large (nearest-neighbor interaction of $\approx 0.5$ eV) and, hence, an essential part of the resulting many-body Hamiltonians. This result confirms previous speculations about the importance of non-local effects in these materials\cite{santoro99, schuwalow10}.
We solve these Hamiltonians within fully self-consistent combined GW and dynamical mean field theory (``GW+DMFT'') \cite{biermann03}, calculating in particular (single particle-) angular resolved photoemission spectra (ARPES) and the (two particle-) charge susceptibility. We identify a clear-cut materials trend starting from Si(111):C deep in a Mott phase to Si(111):Pb which shows tendencies towards metallicity and charge-ordered states driven by non-local interaction terms. Comparing our results to available experimental data yields encouraging insights: \emph{Without adjustable parameters} we reproduce the experimentally measured gap size of insulating Si(111):Sn and its transition to a ``bad-insulator'' at elevated temperatures. Moreover, based on the charge susceptibility, we identify the \emph{electronic} tendency of Si(111):Pb towards charge-ordering of the so-called $3\times 3$ symmetry, which is indeed seen experimentally by scanning tunneling microscopy. Our work is the first one that addresses the electronic properties of real materials on the basis of fully self-consistent GW+DMFT calculations (for a non-self-consistent calculation see \cite{tomczak12}, for self-consistent calculations for models see \cite{sun04, ayral-prl, ayral-prb}). \cite{note1}

\begin{figure}[t]
\includegraphics[width=\columnwidth]{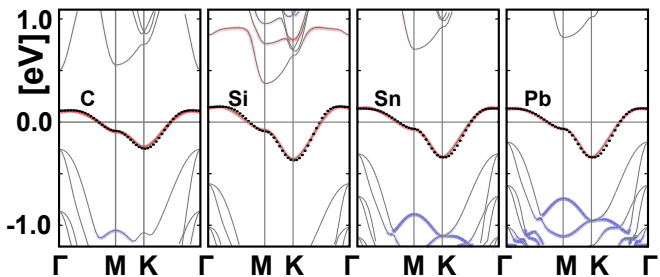}
\caption{(Color online) Bandstructures of the $\alpha$- $\sqrt{3}\times\sqrt{3}$ phases for Si(111):X with X=Sn, Si, C, Pb \cite{silicon}. The color of the bands denotes their respective orbital character. Red color indicates a $p_z$-like ``apical'' character, while the blue color denotes $p_{x,y}$-like (i.e. \emph{planar}) character. The black dots represent the tight binding fit given by Eq.~(\ref{TBHam}) and hopping parameters from Tab.~\ref{uvaluestab}}.
\label{fatbands}
\end{figure}

The single-particle part of the Hamiltonian is calculated in the local density approximation of density functional theory.
In Fig.~\ref{fatbands} we present LDA bandstructures for the series Si(111):$\{{\rm C, Si, Sn, Pb}\}$. For all considered systems the surface-state in the semiconducting gap is indeed responsible for a \emph{well-separated, single band around the Fermi energy}. In red (gray) we plot the contributions stemming from the $p_z$-orbital of the adatom while we plot the adatom $p_{x,y}$-character in blue (dark gray). Even though the actual molecular orbital composition might be complicated, the half-filled surface band has a clear-cut ``apical'' (i.e. carrot-like) character.  
For our calculations presented below we directly use the \emph{ab initio} derived dispersion relation. However, for the purpose of analysis we note that the tight-binding dispersion of the half-filled surface band can be well fitted using up to third-nearest-neighbor hopping ($t$, $t'$, and $t''$) by:
\begin{multline}
\varepsilon_{\bvec{k}}=2t\cdot\left(\cos(k_x)+2\cos(k_x/2)\cos(\sqrt{3}/2 k_y)\right)\\
+2t'\cdot\left(\cos(\sqrt{3}k_y)+2\cos(3k_x/2)\cos(\sqrt{3}/2 k_y)\right)\\
+2t''\cdot\left(\cos(2k_x)+2\cos(k_x)\cos(\sqrt{3} k_y)\right)
\label{TBHam}
\end{multline}
The values for the hopping integrals can be found in Tab.~\ref{uvaluestab} and we plot the analytically calculated bands in Fig.~\ref{fatbands} as the black dashed line. The quality of the fit supports the picture of Wannier-like orbitals with a fast decaying real space overlap on neighboring sites. 

\begin{table}[t]
\caption{\label{uvaluestab} Values of the bare ($V$) and static, screened ($U_0=U(i\nu=0)$) values for on- and intersite nearest neighbor (nn) interaction parameters. Also reported are the values of the static component of the effective $\mathcal{U}(\omega=0)$ calculated from GW+DMFT, see text.}

\begin{tabular*}{\columnwidth}{@{}l*{15}{@{\extracolsep{0pt plus12pt}}l}}
\hline\\[-0.3cm]
 &C&Si&Sn&Pb& \\
\hline\\[-0.2cm]
$t$&$38.0$&$50.0$&$42.0$&$42.0$&[meV]\\
$-t'$&$15.0$&$23.0$&$20.0$&$20.0$&[meV]\\
$t''$&$0.5$&$5.0$&$10.0$&$10.0$&[meV]\\[0.1cm]
\hline\\[-0.2cm]
$U_0$&$1.4$&$1.1$&$1.0$&$0.9$&[eV]\\
$U_1$&$0.5$&$0.5$&$0.5$&$0.5$&[eV]\\
$U_n$& & &$U_1/r_a$& &\\[0.1cm]
\hline\\[-0.2cm]
$V_0$&$6.0$&$4.7$&$4.4$&$4.3$&[eV]\\
$V_1$&$2.8$&$2.8$&$2.7$&$2.8$&[eV]\\[0.1cm]
\hline\\[-0.2cm]
$V_1/\varepsilon^{\rm stat.}_{\rm Si surf.}$&$0.47$&$0.47$&$0.45$&$0.47$&[eV]\\[0.1cm]
\hline\\[-0.2cm]
$\mathcal{U}(\omega=0)$&$1.3$&$0.94$&$0.84$&$0.67 ({\rm ins.})$&[eV]\\
 & & & &$0.54 ({\rm met.})$&[eV]\\[-0.25cm]
\end{tabular*}
\end{table}
In order to determine the interaction parameters as partially screened matrix elements of the Coulomb interaction within the cRPA one has to choose a suitable energy-window around the Fermi energy encompassing the surface band. The bare interaction parameters are calculated by means of explicit evaluation of the radial (Slater-) integrals of the Wannier functions. Subsequently, the dielectric tensor is obtained within cRPA for local and non-local interaction parameters\cite{jpcm12}. The results are summarized in Tab.\ref{uvaluestab}. 

The bare onsite interaction parameters ($V_0$) vary between $6.0$ eV for Si(111):C and $4.3$ eV for Si(111):Pb decreasing monotonously within the series. The onsite $U_0$ is reduced roughly by a factor of $4-5$ due to cRPA screening. At first glance the onsite/static $U_0$ of the order of $1$ eV - about twice the size of the bandwidth - strongly points towards Mott physics. This is, however, a premature conclusion due to the effect of \emph{non-local} interaction terms. The first non-local contribution (nearest-neighbor interaction) $U_1$ [bare $V_1$] is $0.5$ eV [$2.8$ eV]. Remarkably, the value is - opposed to $U_0$ [$V_0$] - almost the same for all materials. The reason is that the intersite overlap of the orbitals is so small that the Coulomb energy corresponds to the electrostatic energy of two point charges. With the virial theorem $\langle E^{\rm tot.} \rangle =1/2 \langle V \rangle$, we quantify this argument by a rescaled hydrogen problem with effective Bohr radius of $6 {\rm \AA}$ ($\approx$distance of adatom sites):
\begin{eqnarray}
\Big \langle \frac{e^2}{r_{\rm rel}} \Big \rangle = \frac{1}{12}
|V_{\rm pot}^{\rm H atom}| = \frac{1}{12} 2 |E_{\rm ground state}^{\rm H atom}|=2.3 eV,
\end{eqnarray}
which roughly matches the value of our bare intersite interaction parameters. The second, likewise remarkable, observation is that the screened values $U_1$ are extremely close to the value we get by assuming a static continuum approximation on the surface of a dielectric medium: $V_1/\varepsilon^{\rm surf}_{\rm Si}$, where $\varepsilon^{\rm surf}_{\rm Si}=\frac{1}{2}(\varepsilon_{\rm Si}+1)$ is the static dielectric constant of silicon on the surface. The reason is straightforward: The adatom distance ($6 {\rm \AA}$) is already large enough compared to the atomic structure of the silicon substrate ($\approx 2{\rm \AA}$) so that \emph{local field effects} (included in cRPA) are negligible. Following this reasoning we can calculate longer range interaction terms by simply scaling $U_1$ with $a/r$, i.e., with the distance in units of the nearest-neighbor distance $a$, i.e., $U_2=U_1/\sqrt{3}$ and so on. In this respect $U_1$ is not only the nearest-neighbor interaction, but the parameter that quantifies the strength of non-local interaction. 

To solve the effective low-energy Hamiltonians resulting from our \emph{parameter-free} downfolding procedure we implement the combined GW+DMFT scheme \cite{biermann03, aryasetiawan03} and calculate spectral properties and charge-charge response functions. Fully self-consistent GW+DMFT was applied to the extended Hubbard model in seminal work by Sun et al. \cite{sun04, note2}, but only recently have numerical techniques for the solution of dynamical impurity models \cite{werner-millis-prl, werner12, casula12} been sufficiently advanced to extract real-frequency information from such calculations \cite{ayral-prl, ayral-prb}. We employ the techniques of the latter two works (in particular a continuous-time quantum Monte Carlo impurity solver within the hybridization expansion \cite{werner-millis-prl}), but implement them for the realistic Hamiltonian derived above. 
Moreover, we go beyond the 'standard' extended Hubbard model and do not restrict the range of the non-local interaction terms. Rather, we include the entire $1/r$ tail by means of an Ewald-type lattice sum.
\begin{figure}[t]%
\includegraphics[width=\columnwidth]{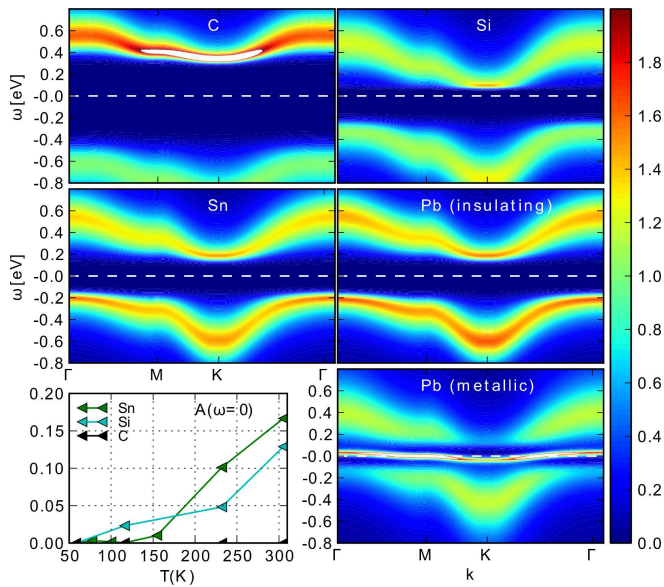}%
\caption{(Color online) Momentum-resolved spectral function  at $T=116$K  of Si(111):X with X=Sn, Si, C, Pb obtained by analytical continuation of GW+DMFT imaginary-time data. The Fermi energy is set to $\varepsilon_{\rm F}=0$ and indicated by the white dashed line. On the bottom right we show the spectral weight at the Fermi energy as a function of temperature.}%
\label{fig02}%
\end{figure}
In Fig.~\ref{fig02} we show momentum-resolved spectral functions from GW+DMFT for all compounds in our series:
As expected from the large onsite interactions compared to the bandwidth we obtain insulating spectra for all four compounds. Interestingly, however, for the Pb compound, in contrast to the other three systems, we find two stable solutions at the temperature of our study ($T=116$K) - one metallic and one insulating. This indicates that we are in a coexistence region of a first order phase transition similar to that seen in the extended Hubbard model\cite{ayral-prb}. 

In all compounds the upper and lower Hubbard bands show substantial dispersion following the bare bandstructure, as expected on general grounds. The insulating gap decreases within the series and we can estimate from the center of mass of the Hubbard bands values of: $1.3$eV (C), $0.8$eV (Si), $0.7$eV (Sn), and $0.5$eV (Pb). However, specifically for the Si(111):\{Sn,Pb\}  we find substantial spectral weight already at $\ge -0.2$eV. Given this small gap, a sizable temperature dependence can be expected.
We have extracted the value of the local (i.e., $\mathbf{k}$-integrated) spectral function at the Fermi-level\cite{note3} (see Fig.~\ref{fig02} bottom left panel). While for Si(111):C the spectral weight transfer to the Fermi energy with temperature is negligible as expected from the spectral function, specifically Si(111):Si and most of all Si(111):Sn display significant transfer of spectral weight on a temperature scale from $50$K to room temperature $300$K. 

Photoemission experiments for Si(111):Sn~\cite{lobo03, cleassen11} (and, possibly\cite{note4}, for Si(111):Pb\cite{dudr04}) observe, indeed, such a temperature dependence and agree well with our results, both, concerning the gap size and temperature scale. Our results - obtained \emph{without any adjustable parameters} - also stand as a theoretical prediction for more extensive studies on Si(111):Pb and the (experimentally so far not studied) Si(111):C compound.
\begin{figure}[t]%
\includegraphics[width=\columnwidth]{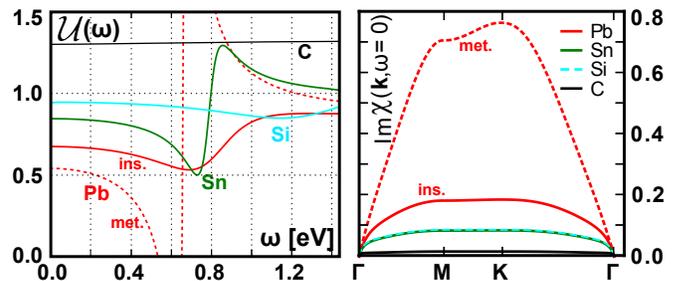}%
\caption{(Color online) Left hand side: Frequency-dependent $\mathcal{U}(\omega)$ (calculated from GW+DMFT) including both, insulating and metallic cases for the Pb system. Right hand side: Imaginary part of the charge-charge susceptibility along the usual path in the Brillouin zone.}%
\label{fig03}%
\end{figure}
Next, we analyze the spectral functions in view of the interaction strengths calculated by cRPA (see Tab.~\ref{uvaluestab}). The gap sizes no longer reflect the energy-scale of the onsite interaction $U_0$ but are reduced due to non-local interactions which \emph{screen} the local interaction by non-local charge fluctuations. 
This physics is naturally present in the GW+DMFT scheme, where non-local effects are incorporated into an effective retarded onsite interaction $\mathcal{U}(\omega)$ (plotted in the left panel of Fig.~\ref{fig03}). The shape of this quantity is reminiscent of screened interactions as calculated, e.g., within the cRPA\cite{aryasetiawan04}, where retardation effects result from downfolding of high-energy degrees of freedom. The GW+DMFT $\mathcal{U}(\omega)$ can be viewed as an effective interaction, where the dynamical character results from downfolding non-local degrees of freedom into a local quantity. 
At large frequencies, screening is not efficient and, hence, $\mathcal{U}(\omega=\infty)=U_0$. On the other hand, the static value $\mathcal{U}(\omega=0)$ can be significantly reduced (up to nearly a factor of 2 for Si(111):Pb). The latter sets the energy scale for the gaps we observe in the spectral function. 
The transition between unscreened high-frequency behavior and the static value takes place at an energy scale $\omega_0$ (plasmonic frequency) characteristic of the non-local charge fluctuations.
The strikingly different behavior of the dynamical effective interactions $\mathcal{U}(\omega)$ reflects the observed materials trend: Si(111):C [Si(111):Si] is [nearly] unaffected by non-local interaction terms and there is barely any screening. The remaining two compounds show, however, large effects. The static values $\mathcal{U}(\omega=0)$ are reduced compared to the onsite interaction to $0.84$eV for Si(111):Sn and to $0.67$eV ($0.54$eV) for the insulating (metallic) solution for Si(111):Pb which leads to the reduced gap sizes. Moreover, plasmonic resonances at energies between $0.6$eV and $0.8$eV stress the importance of non-local interactions/charge-fluctuations for these systems.

\begin{figure}[t]
\includegraphics[width=\columnwidth]{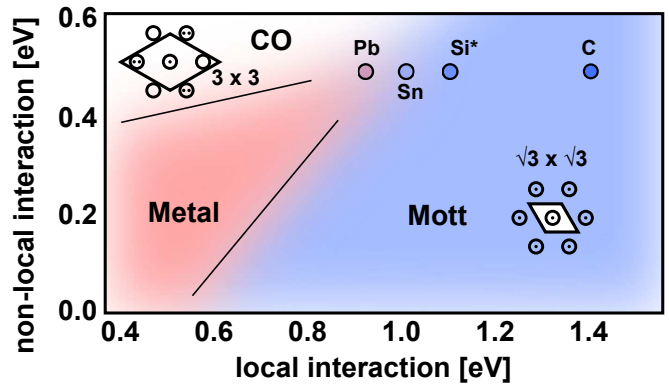}
\caption{(Color online) Schematic local/nonlocal- interaction phase-diagram: The black-bordered circles mark the positions of the adatom systems of our study. Straight lines are guide to the eyes and blurry color indicates coexistence regions. Within the localized (i.e. insulating) phases CO and Mott, small sketches indicate the shape of the surface unit-cell.}%
\label{fig04}
\end{figure}

Besides leading to a retarded, frequency-dependent interaction, the non-local charge fluctuations signal tendencies towards a charge-ordered (CO) state. Analyzing the momentum-dependence of the imaginary part of the charge-charge response function ${\rm Im}\chi(\mathbf{k},\omega=0)$ for the high symmetry points of the Brillouin zone, shown in Fig.~\ref{fig03}, we find for the different materials very particular behavior. The local double occupancy, which corresponds to the integral of the plotted quantity over all momenta, becomes larger towards the end of the series. Most interesting is the case of metallic Si(111):Pb for which we find a distinct structure within the Brillouin zone: The maximum of ${\rm Im}\chi(\mathbf{k},\omega=0)$ at the K symmetry point indicates strong charge fluctuations of the so-called $3\times 3$ symmetry, sketched in Fig,~\ref{fig04}. This order might eventually be frozen in to form a charge ordered ground state which is actually seen in scanning tunneling microscopy for this material\cite{slezak99}. An insulating charge ordered ground state of $3\times 3$ symmetry is, in fact also seen in Ge(111):Sn\cite{avila99} where a concomittant structural distortion (vertical displacement of adatoms) of the same symmetry is seen - our results show, that the instability in the correlated electronic response function is a good candidate for the key player of this feature.

We can summarize our results by drawing a schematic phase-diagram as a function of the strength of local and non-local interactions (represented by the value of $U_1$) as we show in Fig.~\ref{fig04}. For zero non-local interactions our phase diagram describes the Mott-Hubbard metal to insulator transition. The adatom systems are placed at about $0.5$eV of non-local interaction strength. However, due to the difference in the onsite term $U_0$ their respective position in the phase diagram and, hence, their ground state character is different: Si(111):C is deep in the Mott phase with a charge localization defined by one electron per adatom-site.

The Si(111):Si compound\cite{silicon} is also of Mott type with only small values for the double occupancy and little effect of plasmon excitations. However, Si(111):Sn and most dramatic Si(111):Pb (which is actually already in a coexistence region) are much closer to a phase boundary to a metallic phase. Even more peculiar is the obvious tendency of Si(111):Pb towards a charge-ordered phase of $3\times 3$ symmetry indicated by the white region in our phase diagram. 

In conclusion, we have set up a fully self-consistent GW+DMFT scheme for the realistic treatment of correlated surface systems to address the electronic properties of the $\alpha$-phases of adatoms on the Si(111) surface. We reported on the {\it ab initio} construction of the materials-specific low-energy Hamiltonians and, most importantly, on the respective interaction parameters including the long-range Coulomb tail. From these it becomes clear that for the adatom systems taking into account non-local interaction effects is mandatory. We have solved the derived many body Hamiltonians and discussed our finding for momentum-resolved spectral functions, to be compared to ARPES spectra. Without adjustable parameters we reproduced available experimental findings, or, in (most) cases where experiments are missing, made predictions. Specifically, the ARPES spectra for the series, as well as the charge order instabilities in the case of Si(111):Pb are key conclusions/predictions which can provide guidance for further experimental and theoretical studies of semiconductor adatom structures.  

We acknowledge useful discussions with the authors of Ref.~\onlinecite{tournier11} and Ref.~\onlinecite{cleassen11}, as well as with M. Capone, M. Casula, H. Hafermann, H. Jiang, M. Katsnelson, O. Parcollet, and E. Tosatti. This work was supported by the French ANR under project SURMOTT and IDRIS/GENCI under project 129313.


\begin{thebibliography}{10}

\bibitem{tosatti74}E. Tosatti and P. W. Anderson. {Japanese Journal of Applied Physics} {\bf 2S2},381-388 (1974)

\bibitem{uhrberg85}R. I. G. Uhrberg, G. V. Hansson, J. M. Nicholls, P. E. S. Persson and S. A. Flodstr\"om. {Phys. Rev. B} {\bf 31},3805 (1985)

\bibitem{grehk93}T. M. Grehk, L. S. O. Johansson, U. O. Karlsson and A. S. Fl\"odstrom. {Phys. Rev. B} {\bf 47},13887 (1993)

\bibitem{weitering93}H. H. Weitering, J. Chen, N. J. DiNardo and E. W. Plummer. {Phys. Rev. B} {\bf 48},8119 (1993)

\bibitem{carpinelli96}J. M. Carpinelli, H. H. Weitering, E. W. Plummer and R. Stumpf. {Nature} {\bf 381},398 (1996)

\bibitem{carpinelli97}J. M. Carpinelli, H. H. Weitering, M. Bartkowiak, R. Stumpf and E. W. Plummer. {Phys. Rev. Lett.} {\bf 79},2859 (1997)

\bibitem{weitering97}H. H. Weitering, X. Shi, P. D. Johnson, J. Chen, N. J. DiNardo and K. Kempa. {Phys. Rev. Lett.} {\bf 78},1331 (1997)

\bibitem{slezak99} J. Slez\'ak, P. Mutombo, and V. Ch\'ab  Phys. Rev. B {\bf 60}, 13328 (1999)

\bibitem{lay01}G. L. Lay, M. G. Rad, M. Gthelid, U. Karlsson, J. Avila and M. Asensio. {Applied Surface Science} {\bf 175-176},201 (2001)

\bibitem{lobo03} J. Lobo, A. Tejeda, A. Mugarza, and E. G. Michel, Phys. Rev. B {\bf 68}, 235332 (2003)

\bibitem{dudr04} V. Dudr, N. Tsud, S. Fab\'ik, B. Ressel, M. Vondr\'acek, K.C. Prince, V. Matol\'in, V. Ch\'ab, Surface Science 566, {\bf 804} (2004)

\bibitem{pignedoli04}C. A. Pignedoli, A. Catellani, P. Castrucci, A. Sgarlata, M. Scarselli, M. De Crescenzi and C. M. Bertoni. {Phys. Rev. B} {\bf 69},113313 (2004)

\bibitem{upton05}M. H. Upton, T. Miller and T.-C. Chiang. {Phys. Rev. B} {\bf 71},033403 (2005)

\bibitem{modesti07} S. Modesti, L. Petaccia, G. Ceballos, I. Vobornik, G. Panaccione, G. Rossi, L. Ottaviano, R. Larciprete, S. Lizzit, and A. Goldoni, Phys. Rev. Lett. {\bf 98}, 126401 (2007) 

\bibitem{cardenas09}L. A. Cardenas, Y. Fagot-Revurat, L. Moreau, B. Kierren and D. Malterre. {Phys. Rev. Lett.} {\bf 103},046804 (2009)

\bibitem{zhang10}T. Zhang, P. Cheng, W.-J. Li, Y.-J. Sun, G. Wang, X.-G. Zhu, K. He, L. Wang, X. Ma, X. Chen, Y. Wang, Y. Liu, H.-Q. Lin, J.-F. Jia and Q.-K. Xue. {Nat Phys} {\bf 6},104 (2010)

\bibitem{tournier11}C. Tournier-Colletta, L. Cardenas, Y. Fagot-Revurat, A. Tejeda, B. Kierren and D. Malterre. {Phys. Rev. B} {\bf 84},155443 (2011)

\bibitem{cleassen11} G. Li, P. H\"opfner, J. Sch\"afer, C. Blumenstein, S. Meyer, A. Bostwick, E. Rotenberg, R. Claessen, and W. Hanke, arXiv:1112.5062 (2011)

\bibitem{kaxiras90}E. Kaxiras, K. C. Pandey, F. J. Himpsel and R. M. Tromp. {Phys. Rev. B} {\bf 41},1262 (1990)

\bibitem{brommer92}K. D. Brommer, M. Needels, B. E. Larson and J. D. Joannopoulos. {Phys. Rev. Lett.} {\bf 68},1355 (1992)

\bibitem{santoro99}G. Santoro, S. Scandolo and E. Tosatti. {Phys. Rev. B} {\bf 59},1891--1901 (1999)

\bibitem{hellberg99}C. S. Hellberg and S. C. Erwin. {Phys. Rev. Lett.} {\bf 83},1003--1006 (1999)

\bibitem{aizawa99}H. Aizawa, M. Tsukada, N. Sato and S. Hasegawa. {Surface Science} {\bf 429},L509--L514 (1999)

\bibitem{profeta00}G. Profeta, A. Continenza, L. Ottaviano, W. Mannstadt and A. J. Freeman. {Phys. Rev. B} {\bf 62},1556 (2000)

\bibitem{shi02}H. Q. Shi, M. W. Radny and P. V. Smith. {Phys. Rev. B} {\bf 66},085329 (2002)

\bibitem{shi04}H. Q. Shi, M. W. Radny and P. V. Smith. {Phys. Rev. B} {\bf 70},235325 (2004)

\bibitem{profeta05}G. Profeta and E. Tosatti. {Phys. Rev. Lett.} {\bf 95},206801 (2005)

\bibitem{profeta07}G. Profeta and E. Tosatti. {Phys. Rev. Lett.} {\bf 98},086401 (2007)

\bibitem{schuwalow10}S. Schuwalow, D. Grieger and F. Lechermann. {Phys. Rev. B} {\bf 82},035116 (2010)

\bibitem{chaput11}L. Chaput, C. Tournier-Colletta, L. Cardenas, A. Tejeda, B. Kierren, D. Malterre, Y. Fagot-Revurat, P. Le F\`evre, F. Bertran, A. Taleb-Ibrahimi, D. G. Trabada, J. Ortega and F. Flores. {Phys. Rev. Lett.} {\bf 107},187603 (2011)

\bibitem{li11}G. Li, M. Laubach, A. Fleszar and W. Hanke. {Phys. Rev. B} {\bf 83},041104 (2011)

\bibitem{aryasetiawan04} F. Aryasetiawan, M. Imada, A. Georges, G. Kotliar, S. Biermann, and A. I. Lichtenstein, Phys. Rev. B {\bf 70}, 195104 (2004)

\bibitem{vaugier12}L. Vaugier, H. Jiang and S. Biermann, Phys. Rev. B {\bf 86}, 165105 (2012)

\bibitem{jpcm12} P. Hansmann, L. Vaugier, H. Jiang and S. Biermann. {(to be published in JPCM in press)}

\bibitem{biermann03} S. Biermann, F. Aryasetiawan, and A. Georges, Phys. Rev. Lett. {\bf 90}, 086402 (2003)

\bibitem{tomczak12} J. Tomczak, M. Casula, T. Miyake, F. Aryasetiawan, and S. Biermann, Europhysics Letters, {\bf 100}, 67001 (2012)

\bibitem{sun04} P. Sun and G. Kotliar, Phys. Rev. Lett. {\bf 92}, 196402 (2004)

\bibitem{ayral-prl} T. Ayral, P. Werner, and S. Biermann, Phys. Rev. Lett. {\bf 109}, 226401 (2012)

\bibitem{ayral-prb} T. Ayral, S. Biermann , and P. Werner, arXiv:1210.2712 (2012)

\bibitem{karlssonJPCM} K. Karlsson, J. Phys.: Condens. Matter {\bf 17} 7573 (2005)

\bibitem{tarantoArxiv12} C. Taranto, M. Kaltak, N. Parragh, G. Sangiovanni, G. Kresse, A. Toschi, K. Held, http://arxiv.org/abs/1211.1324 (2012)

\bibitem{aryasetiawan03}  F. Aryasetiawan, S. Biermann, and A. Georges, http://arxiv.org/abs/cond-mat/0401653 (2004)

\bibitem{sun02} P. Sun and G. Kotliar, Phys. Rev. B {\bf 66}, 085120 (2002)

\bibitem{werner-millis-prl} P. Werner and A. J. Millis, Phys. Rev. Lett. {\bf 104}, 146401 (2010)

\bibitem{werner12} P. Werner, M. Casula, T. Miyake, F. Aryasetiawan, A. Millis. S. Biermann, {Nature Physics} {\bf 8}, 331 (2012) 

\bibitem{casula12} M. Casula, A. Rubtsov, S. Biermann {Phys. Rev. B} {\bf 85}, 035115 (2012)

\bibitem{avila99} J. Avila, A. Mascaraque, E. G. Michel, M. C. Asensio, G. LeLay, J. Ortega, R. Perez, and F. Flores, Phys. Rev. Lett. {\bf 82}, 442 (1999)

\bibitem{note1} We also mention attempts of simplified combined schemes using static screened interactions in \cite{biermann03, karlssonJPCM, tarantoArxiv12}.

\bibitem{note2} See also \cite{sun02} for a variant of the GW+DMFT scheme.

\bibitem{note3} From the imaginary-time Greens function at $G(\tau=\beta/2)$

\bibitem{note4} It is experimentally difficult to determine the symmetry of the Si(111):Pb ground state - this challange can, in fact, be explained by our results of charge order instabilities.

\bibitem{silicon} The structure Si(111):Si is calculated/presented as a hypothetical structure to complete the series - it is not stable in the $\sqrt{3}\times\sqrt{3}$ phase.

\end{thebibliography}
\end{document}